\documentclass[fleqn,twoside]{article}
\usepackage[headings]{espcrc2}

\readRCS
$Id: espcrc2.tex,v 1.2 2004/02/24 11:22:11 spepping Exp $
\usepackage{graphicx}
\usepackage{epsfig}
\usepackage[figuresright]{rotating}

\newcommand{\AmS}{{\protect\the\textfont2
  A\kern-.1667em\lower.5ex\hbox{M}\kern-.125emS}}

\title{TOTEM: The experiment to measure the total proton--proton
cross section at LHC}

\author{S. Lami\address[INFN]{Istituto Nazionale di Fisica Nucleare, Sezione di Pisa, \\ 
        Largo B. Pontecorvo 3, I--56127 Pisa, Italy}%
        \thanks{For the TOTEM Collaboration}}
       
\runtitle{TOTEM: The experiment to measure the total proton--proton cross section at LHC}
\runauthor{S. Lami}

\begin{document}

\begin{abstract}
The current large uncertainty on the extrapolation
of the proton--proton  total cross section at the LHC
energy will be resolved by the precise measurement
by the TOTEM experiment.
Its accurate studies on the basic properties of proton--proton
collisions at the maximum accelerator energy could
provide a significant contribution to the
understanding of cosmic ray physics.
\vspace{1pc}
\end{abstract}

\maketitle

\section{INTRODUCTION}

TOTEM is an experiment dedicated to the measurement of the proton--proton
total cross section ($\sigma_{TOT}^{pp}$) at LHC, i.e. the probability that two protons interact
at the center of mass energy of 14 TeV. In addition to that, it will
also study the proton--proton elastic scattering and  diffractive dissociation
processes.

TOTEM foresees specific measurements and experimental techniques which are
very different from the other `general purpose' experiments at LHC.
A precise `luminosity independent' measurement of $\sigma_{TOT}^{pp}$ will
be achievable in special beam optics runs by simultaneously measuring: 1) the 
elastic scattering rate at low transfer momentum, possibly as small as $t=10^{-3}$ GeV$^2$, 
and 2) the inelastic scattering rate with the largest possible coverage to
reduce losses to few percents.
The first goal requires detectors located into units mounted into the vacuum chamber of
the accelerator, called Roman Pots (RPs), as the scattered protons are emitted at angles
of the order of 10~$\mu$rad, therefore without leaving the beam--pipe.
The latter requires the measurement of all the inelastically produced particles
in the very forward direction with respect to the $pp$ collision point; this can
be achieved by using tracking detector telescopes with a complete azimuthal
coverage around the beam--pipe.

In the following, a brief description of the experimental apparatus
is given. More details can be found in \cite{TDR}. I will then
discuss physics issues such as the measurement of  $\sigma_{TOT}^{pp}$
and the validation of hadronic shower models used in cosmic ray physics.

\section{EXPERIMENTAL APPARATUS}

\begin{figure}[ht!]
\vspace*{-6pt}
\hspace*{-12pt}\epsfig{file=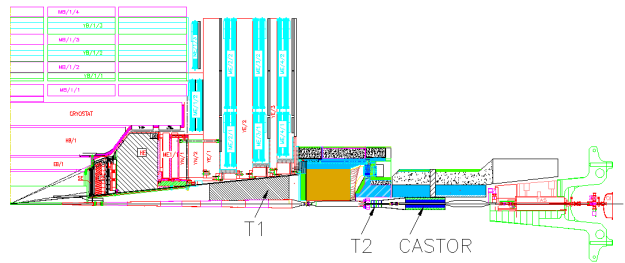,width=1.15\linewidth,angle=0}
\vspace*{-27pt}
\caption{The TOTEM forward trackers T1 and T2 into the forward region of the CMS detector.}
\label{fig:cmsbello}
\end{figure}

The experimental setup comprises Roman Pot detectors to measure the leading protons
at $\pm$~147 and 220 meters from the LHC interaction point IP5, while the
inelastic telescopes T1 and T2 are located inside the CMS end--caps around the
beam--pipe, as shown in Fig.~\ref{fig:cmsbello}.
 Note also the planned forward calorimeter
Castor, under CMS's responsibility, with the same acceptance of the T2 detector.
Rapidity gaps and forward particle flows could be measured by the TOTEM telescopes T1 and T2,
while forward energy flows could be measured by T2 and the CMS Castor forward calorimeter.
\begin{figure}[htb]
\hspace*{-12pt}\epsfig{file=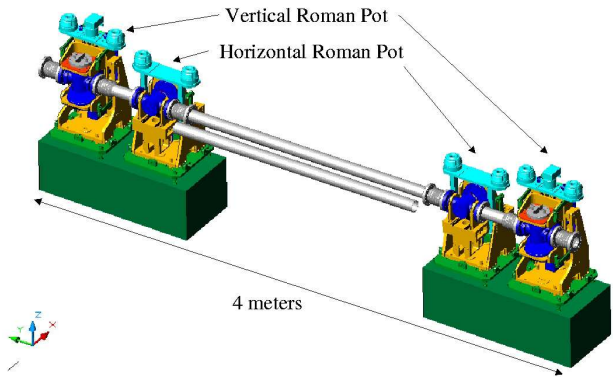,width=1.12\linewidth,angle=0}
\vspace*{-27pt}
\caption{The TOTEM Roman Pot station.}
\label{fig:station}
\end{figure}

\begin{figure}[htb]
\centering{\epsfig{file=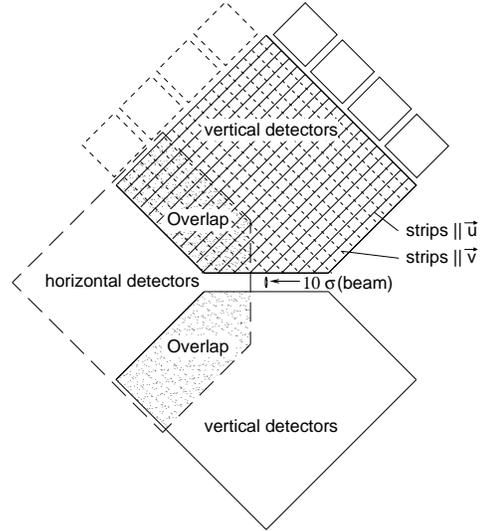,width=0.83\linewidth,angle=0}}
\vspace*{-21pt}
\caption{Arrangement of the detectors in the two vertical and the one horizontal RPs of a station.}
\label{fig:overlap}
\end{figure}

Each RP station has two units 4 m apart, as shown in Fig.~\ref{fig:station}. 
Each unit has two vertical pots approaching the
beam from the top and the bottom, where beams are
usually more stable, and one lateral pot sensitive to 
diffractive protons. Furthermore, the overlap
between the horizontal and the vertical pots (Fig.~\ref{fig:overlap}) will 
serve for measuring the relative distance of the vertical detectors.
Each pot will contain 5+5 planes of Silicon detectors, their strips having 
orientations of $\pm 45^{\rm o}$ w.r.t. the detector edge, and a pitch
of 66 $\mu$m.\\
In order to optimize the measurement of microscopic proton scattering angles,
the RP detector edge has to move as close to the beam as $\sim$1~mm and, therefore,
the edge dead area has to be greatly minimized. A new edgeless technology of Silicon
microstrips has been developed, where a current terminating structure will reduce
to only 50 $\mu$m the decoupling area between edge and sensitive volume.
Strong magnetic dipoles between the RP stations will provide a powerful
magnetic spectrometer. Particle momenta will be measured with an accuracy
of a few parts per thousand, allowing an accurate determination of the
momentum loss of quasi--elastically scattered protons in diffractive processes.

The T1 telescopes on both sides of the interaction point will cover the 
pseudorapidity range $3.1 < |\eta| < 4.8$. It will consist of 
five planes, each composed of six trapezoidal Cathode Strip Chambers (CSC). 
Each detector will measure three projections: 
one set of anode wires with a pitch of 3\,mm measuring the radial coordinate
and two sets of cathode strips with a pitch of 5\,mm, rotated
by $\pm 60^{\rm o}$ with respect to the wires. 
The radial measurement will provide level-1 trigger information and will
be used for vertex reconstruction in order to suppress beam-gas background. 
Beam tests of final prototypes have shown a spatial resolution of 0.36\,mm
in the radial and 0.62\,mm in the azimuthal coordinate.

\begin{figure}[htb]
\centering{\epsfig{file=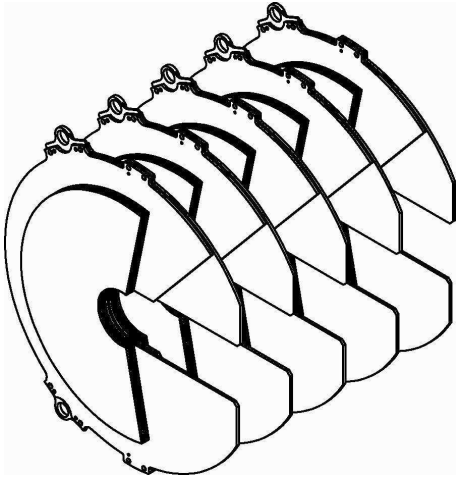,width=0.8\linewidth,angle=0}}
\vspace*{-23pt}
\caption{One half--arm of the TOTEM T2 telescope.}
\label{fig:t2}
\vspace{-19pt}
\end{figure}
For T2, which extends the acceptance into the range $5.2 < |\eta| < 6.7$, the
Gas Electron Multiplier (GEM) technology as used successfully in 
COMPASS~\cite{compassgem} has been chosen. 
GEMs are
gas-filled detectors in which  the charge
amplification structure is decoupled from the charge collection and 
readout structure.
Furthermore, they combine good spatial resolution with very high rate
capability and a good resistance to radiation. The T2 telescope will be
placed 13.5\,m away from the IP5 and the GEMs employed will have an almost
semicircular shape, with an inner radius matching the beam pipe. Each half--arm
of T2 will have a set of 10 aligned detector planes mounted `back-to-back' on each side
of the vacuum pipe (Fig.~\ref{fig:t2}). To avoid efficiency losses, the angular coverage of
each half plane is more than 180$^{\circ}$.
The read-out boards will have
two separate layers with different patterns: one with 256 concentric circular
strips, 80\,$\mu$m wide and with a pitch of 400\,$\mu$m, and the other with 
a matrix of pads varying in size from $2 \times 2\,\rm mm^{2}$ to 
$7 \times 7\,\rm mm^{2}$ (for a constant 
$\Delta \eta \times \Delta \phi = 0.06 \times 0.017 \pi$).
The pad information will also provide level-1 trigger information.
A final prototype has been successfully tested in the 2004 test-beam.

The read-out of all TOTEM detectors will be based on the digital VFAT chips, 
enhancing the system uniformity from the point of view of the data processing chain.

\section{PHYSICS GOALS}

\begin{figure}[htb]
\hspace*{-9pt}\epsfig{file=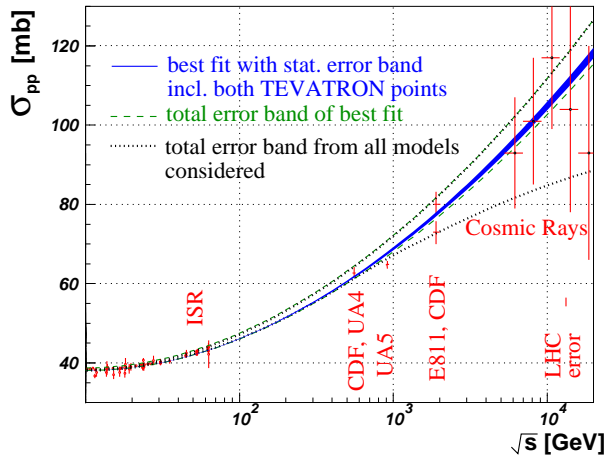,width=1.09\linewidth,angle=0}
\vspace*{-35pt}
\caption{Predictions for total $pp$ cross sections~\cite{compete}, including ISR and cosmic ray data.}
\label{fig:sigmapp}
\end{figure}

Fig.~\ref{fig:sigmapp} shows recent predictions~\cite{compete} for the energy
dependence of the total $pp$ cross section $\sigma_{TOT}^{pp}$ by fitting
all available data.
The black error band shows the statistical errors to the best fit, the closest
curves near it give the sum of statistical and systematic errors to the best fit
due to the ambiguity in the Tevatron data, and the highest and lowest curves show
the total error bands from all models considered. For the LHC energy a value
 $\sigma_{TOT}^{pp}= 111.5 \pm 1.2 ^{+4.1}_{-2.1} $ mb is obtained from the best fit,
while the total error band ranges in the 90--130 mb interval.
This large theoretical uncertainty is due to the current lack of a fully satisfactory
theoretical explanation of the cross section in low momentum transfer collisions, 
and their description relies on phenomenological models.
TOTEM aims to measure $\sigma_{TOT}^{pp}$ with a precision of 1\% or 1 mb, therefore
discriminating among the different models.

The typical instantaneous luminosity for the TOTEM  $\sigma_{TOT}^{pp}$
measurement will be of the order of 10$^{28} cm^{-2}s^{-1}$. This is
due to the high machine optics $\beta^*$ value - 1540 m for the 1\% measurement -
required in special runs in order to keep as small as possible
 the  beam angular divergence for
a precise measurement of small scattering angles.
As a consequence, the beam size at the interaction point increases.
Therefore, in order to avoid extra interactions between the
colliding beams inside the common vacuum chamber, a small
number of bunches as well as a zero crossing angle are desirable,
resulting in the forementioned luminosity for an optics with
 $\beta^* =$ 1540 m and 43 bunches.

A special beam optics  with $\beta^* =$ 90 m (and a luminosity close to 
10$^{30} cm^{-2}s^{-1}$), still enabling
a  $\sigma_{TOT}^{pp}$ measurement with a few percent uncertainty,
would also provide an excellent measurement of the momentum loss
of diffractive protons, opening the studies of soft and semihard
diffraction, the latter in combination with the CMS detectors.

Without an accurate measurement of the machine luminosity, the
only practical way to determine  $\sigma_{TOT}^{pp}$ is the
`luminosity independent' method which combines the total rate equation,
as the sum of elastic and inelastic interactions, to the optical theorem
relation between  $\sigma_{TOT}^{pp}$, luminosity and the imaginary part
of the forward amplitude, such that the luminosity is eliminated
and $\sigma_{TOT}^{pp}$ can be written as a function of measurable
rates:
\begin{equation}
{\sigma_{TOT}^{pp}} = {16\pi  \over{(1+\rho^2)}} {(dN_{el}/dt)_{t=0} \over{(N_{el}+N_{inel})}} 
\end{equation}
where the optical point at $t=0$ has to be extrapolated from the measurement of the
elastic scattering at low momentum transfers.

Let's consider the measurement with the special  $\beta^* =$ 1540 m optics.
The statistical error on the extrapolation of the elastic cross section
 at $t=0$ is less than 0.1\% already after 10 hours of data taking.
The systematic error is dominated by the insufficient knowledge at very low $t$ of the
functional form for the extrapolation, and it will be less than 0.5\% if
angles as low as 14$\mu$rad - equivalent to about $ t = 10^{-2}$ GeV$^2$ - could
be measured, well within the experiment expectations.
Elastic events will be selected by a double--arm trigger, with a signal
from left amd right RPs, plus the collinearity of the two protons.
The vertex reconstruction will help eliminating the background
from beam--gas and beam halo events. In addition to that, the selection
of inelastic events will include a single--arm trigger in coincidence
with a leading proton in the opposite side RP for the single diffractive
events. Single diffractive events with masses below 10 GeV will fail
the trigger, but the resulting loss can be corrected. This is estimated
to give the largest contribution to the total error on the total rate
measurement which is about 0.8\%. The sum of all uncertainties results
in  an error of about 1\% on  $\sigma_{TOT}^{pp}$.

\begin{figure}[htb]
\vspace*{-8pt}
\hspace*{-10pt}\epsfig{file=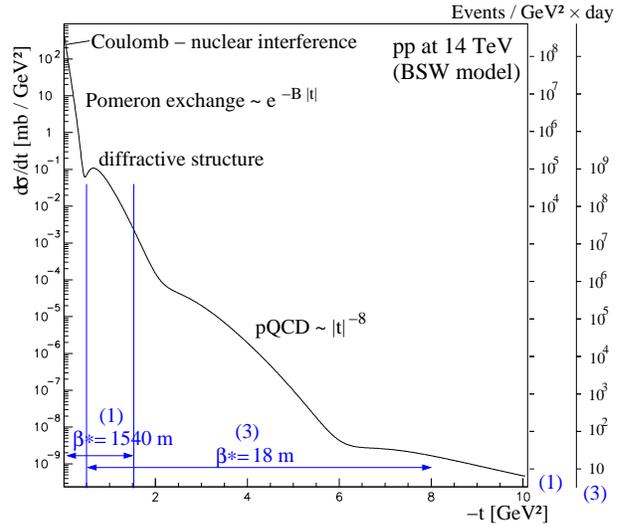,width=1.15\linewidth,angle=0}
\vspace*{-35pt}\caption{Elastic scattering cross section, using the model by BSW~\cite{bsw}.
The number of events at the right side of the plot refers to a day running
at two different optics configurations.}
\label{fig:sigmael}
\end{figure}
An important measurement to understand the mechanism of high energy collisions
is the ratio of the elastic over total cross sections, which in the past
was found to increase with energy at CERN and Fermilab. Among several phenomenological models,
the one by Bourelly, Soffer and Wu~\cite{bsw} foresees for instance that, at very high
energies, the effective interaction radius of the colliding hadrons increase
as log$s$,  $\sigma_{TOT}$ increase as log$s^2$ and   $\sigma_{el}/\sigma_{TOT}$
approach 1/2. At LHC energy, the elastic cross section is supposed to be about 30~mb.
To discriminate between different models it is thus important to precisely
measure the elastic scattering over the largest possible $t$ region.
The $t$ distribution assuming the BSW model is given in Fig.~\ref{fig:sigmael}.
It extends over 11 orders of magnitude and has therefore to be measured
with different optics settings. 
The exponential fall at low $t$ is followed by a diffractive structure
at $\sim$1 GeV$^2$ and continues to large $t$-values where perturbative
calculations suggest a power--law behaviour ($t^{-8}$).
The number of events at the right side of the plot refers to a day running
at two conditions (optics with special  $\beta^*=$ 1540 m and injection
 $\beta^*=$ 18 m, respectively). The maximal detectable $t$-value due
to aperture limitations in the LHC is 8 GeV$^2$ with  $\beta^*=$ 18~m.

While the measurement of the total cross section and the elastic scattering
can be performed using only TOTEM detectors, the integration of TOTEM
into the general purpose detector CMS offers the prospect
of more detailed studies of diffractive events. The TOTEM triggers, combining information
from the inelastic detectors and the silicon detectors in the RPs, can be incorporated
into the general CMS trigger scheme. The CMS experiment extended by the TOTEM
detectors provides the largest acceptance detector ever implemented at 
a hadron collider.

\section{THE VHE COSMIC RAYS CONNECTION}

The aim of the TOTEM experiment is to obtain accurate information on the basic properties
of proton--proton collisions at the maximum accelerator energy, thus providing
a significant contribution to the understanding of very high energy cosmic ray physics.

Primary cosmic rays in the PeV (10$^{15}$ eV) energy range and above are a challenging
issue in astrophysics.
The LHC center of mass energy corresponds to a 100 PeV energy for a fixed target
collision in the air, at the same time providing a high event rate relative to
the very low rate of cosmic particles in this energy domain.

A primary cosmic ray entering the upper atmosphere experiences a nuclear interaction,
with the production of nuclear fragments and $\pi$ mesons, starting an air shower
with hadronic, electromagnetic and muon components. The real challenge is to 
determine the nature of the primary interaction and the energy and composition
of the incident particle from the measurement of the shower.
Several high energy hadronic interaction models are nowadays available, which predict
energy flow, multiplicity and other quantities of such showers.
There are large differences between the predictions of currently available
models, with significant inconsistencies in the forward region.

Among the several quantities that can be measured by TOTEM and CMS, and compared with
model predictions, are: energy flow, transverse energy,
elastic/total cross section, fraction of diffractive events, particle
multiplicity, ratio of the number of hadronic secondaries to that of
leptonic secondaries, and the distribution of the inelasticity coefficient of the
incident nucleon (i.e. the ratio of the energy of the most energetic
outgoing particle to the energy of the incident particle, it defines the
shape of the shower).

Samples of events obtained with some of the available generators (QGSjet
0.1~\cite{Kalmykov:1997te}, SIBYLL 2.1~\cite{Engel:1999db}, DPMJet 
3~\cite{Roesler:2001mn}, neXus 3~\cite{Kalmykov:1997te}) were passed
through the simulation of T1, T2 and Castor detectors. 
Two data samples
were considered: all inelastic collisions and diffractive events.
Diffractive events were defined as those with a leading proton with
a momentum loss $0.003 < \xi < 0.05$.

\begin{figure*}[htb]
\newcommand{\cc}[1]{\multicolumn{1}{c}{#1}}
\renewcommand{\tabcolsep}{1pc} 
\renewcommand{\arraystretch}{1.0} 
\centering{\epsfig{file=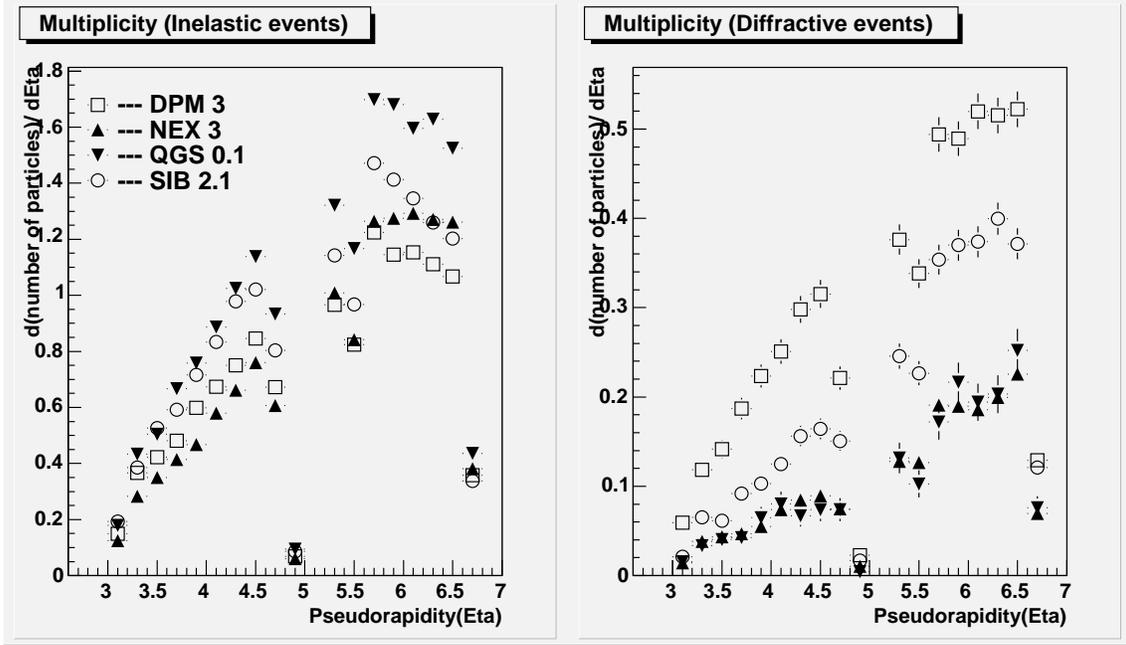,width=0.54\linewidth,angle=270}}
\vspace*{-20pt}
\caption{
Charged particle multiplicity as a function of pseudorapidity for
inelastic events (left) and diffractive events (right).
\label{fig:cosmic-multiplicity}}
\end{figure*}

As an example, Figure~\ref{fig:cosmic-multiplicity} shows the predictions
of the quoted Monte Carlo generators for the charged
particle multiplicity. 
Table~\ref{tab:cosmic-fraction-castor} shows
the fraction of diffractive events expected in Castor.  Significant
differences in the predictions are evident. The differences are larger
for the diffractive events rather than for inelastic events.  Appreciable
differences are also observed for the inelasticity coefficient as well as
for the energy flow. The study of the features of
diffractive and inelastic events as measured in Castor and TOTEM may thus
be used to validate/tune the generators~\cite{ptdr}.

\section{CONCLUSIONS}
TOTEM will be ready for data taking at LHC start.
The undergoing production of the final TOTEM detectors is
proceeding fine and, at the time of writing, a 
fraction of them is under test on the CERN SPS H8 beam line.
The RPs are foreseen to be installed in Spring 2007, while
all the detectors will be ready by July 2007.

TOTEM precise measurements of basic properties of proton--proton
collisions at LHC will
provide a significant contribution to the
understanding of cosmic ray physics,
by discriminating among currently popular shower
models.

\begin{table}[htb]
\caption{Fraction of diffractive events in the Castor pseudorapidity 
acceptance.}
\vspace*{5pt}
\centering
\small {
\begin{tabular}{|c|c|c|c|}
\hline
 QGS-01 & SIB-2.1 &  DPM-3    & NEXus3   \\
\hline
\hline
 4.5\%  & 12.4\%  & 13.6\%    & 24.3\% \\
\hline
\end{tabular}
}
\label{tab:cosmic-fraction-castor}
\end{table}
%
%


\begin{thebibliography}{9}
\bibitem{TDR} TOTEM: Letter of Intent, CERN-LHCC 97-49;
Technical Proposal, CERN-LHCC 99-7;
Technical Design Report, CERN-LHCC-2004-002;
and references therein.
\bibitem{compassgem} C. Altunbas et al., `Construction, test and commissioning
of the triple-GEM tracking detector for Compass', NIM A 490 (2002) 177.
\bibitem{compete} J.R. Codell et al., Phys. Rev. Lett. 89, 201801 (2002).
\bibitem{bsw} C. Bourelly, J. Soffer and T.T. Wu, Eur. Phys. J. C28 (2003) 97.
\bibitem{Kalmykov:1997te} N.N. Kalmykov, S. Ostapchenko, A. Pavlov, 
Nucl. Phys. Proc. Suppl. 52B (1997) 17.
\bibitem{Engel:1999db} R. Engel, T.K. Gaisser, T. Stanev and P. Lipari,
`Air shower calculations with the new version of SIBYLL',
Prepared for 26th International Cosmic Ray Conference (ICRC99), Salt Lake City, Utah, 17-25 Aug 1999.
\bibitem{Roesler:2001mn} S. Roesler, R. Engel and J. Ranft,
`The event generator DPMJET-III at cosmic ray energies',
Prepared for 27th International Cosmic Ray Conference (ICRC2001), Hamburg, Germany, 7-15 Aug 2001.
\bibitem{ptdr} The CMS and TOTEM diffractive and forward physics working group,
`Prospects for Diffractive and Forward Physics at the LHC', in preparation.
\end{thebibliography}
\end{document}